\documentclass[twocolumn]{article}
\usepackage{graphics,cite,amssymb,epsfig,float,psfrag,dcolumn}
\newcommand{\beq}{\begin{equation}}
\newcommand{\eequ}{\end{equation}}
\newcommand{\eeq}{\end{equation}}
\def\bea{\begin{eqnarray}}
\def\eea{\end{eqnarray}}
\def\as{\relax\ifmmode\alpha_s\else{$\alpha_s${ }}\fi}
\def \pt{\relax\ifmmode{p_t}\else{$p_t${ }}\fi}
\def\nn{\nonumber}
\newcommand{\noi}{\noindent}

\def\gl{\Gamma}

\def\l{\ell}

\def\d{{\rm d}}

\def\sh{\hat{s}}

\def\a{{\cal A}}

\def\uh{{\hat{u}}}

\def\as{{\alpha_s}}
\def\be{\begin{equation}}
\def\ee{\end{equation}}
\def\ba{\begin{eqnarray}}
\def\ea{\end{eqnarray}}

\newif\ifdtup

\def\eqal2#1{\,\vcenter{\openup1\jot
\caja   \ialign{\strut \hfil$\displaystyle{##}$&\hfil$
\displaystyle{{}##}$\hfil &$
\displaystyle{{}##}$\hfil\crcr#1\crcr}}\,}

\begin{document}

\title{
\begin{flushright}\normalsize
\vspace{0cm}
DESY 02-143 \\ 
hep-ph/0209191\\
September 2002
\vspace{1.cm}
\end{flushright}
\bf  
$B \to K^* \ell^+ \ell^-\big (\rho~ \ell \nu_{\ell}\big)$ helicity
analysis in the LEET\footnote{Talk presented at the $10^{th}$
International Conference on Supersymmetry and Unification of Fundamental
Interactions (SUSY02), Hamburg, Germany, 10-23 June 2002.
Based on work together with Ahmed Ali~\cite{Ali:2002qc}.}
\unboldmath}
\author{
	A. Salim Safir \\
        Deutsches Elektronen-Synchrotron, DESY, D-22603 Hamburg, Germany \\
        E-mail~: safir@mail.desy.de} 

\par \maketitle
\maketitle
\vspace{0.2truecm}

\begin{abstract}
We calculate the independent helicity amplitudes in the decays $B \to K^*
\ell^+ \ell^-$ and $B \to \rho \ell \nu_\ell$ in the so-called
Large-Energy-Effective-Theory (LEET). Taking into account the dominant
$O(\alpha_s)$ and $SU(3)$ symmetry-breaking effects, we calculate 
various single (and total) distributions in these decays making use of
the presently available data and decay form factors calculated in the QCD sum
rule approach. Differential decay rates in the dilepton 
invariant mass and the Forward-Backward asymmetry in $B \to K^* \ell^+ \ell^-$
are worked out. Measurements of the ratios $R_i(s) \equiv d
\Gamma_{H_i}(s)(B \to K^* \ell^+ \ell^-)/ d \Gamma_{H_i}(s)(B \to \rho
\ell \nu_\ell)$, involving the helicity amplitudes $H_i(s)$, $i=0,+1,
-1$, as precision tests of 
the standard model in semileptonic rare $B$-decays are emphasized. We
argue that $R_0(s)$ and $R_{-}(s)$ can be used to determine the CKM ratio
$\vert V_{ub}\vert/\vert V_{ts} \vert$ and search for new physics, where
the later is illustrated by supersymmetry.
\end{abstract}
\par \maketitle

%----------------------------------------
\section{Introduction}\hspace*{\parindent}
Rare $B$ decays involving flavour-changing-neutral-current (FCNC) 
transitions, such as $b \to s \gamma$ and $b \to s \ell^+ \ell^-$, have
received a lot of theoretical interest \cite{Greub:1999sv}, especially
after the first measurements reported by the
CLEO collaboration \cite{Alam:1995aw} of the $B \to X_s \gamma$
decay. The current world average based on the improved measurements by
the CLEO \cite{Chen:2001fj}, ALEPH \cite{alephbsg}
and BELLE \cite{bellebsg} collaborations, 
${\cal B}(B \to X_s \gamma)=(3.22 \pm 0.40) \times 10^{-4}$, 
is in good agreement with the estimates of the standard model (SM) 
\cite{Chetyrkin:1997vx,Kagan:1999ym,Gambino:2001ew}, which we shall take
as ${\cal B}(B \to X_s \gamma)=(3.50 \pm 0.50) \times 10^{-4}$, reflecting
the parametric uncertainties dominated by the scheme-dependence of the
quark masses. The decay $B \to X_s \gamma$ also provides useful 
constraints on the parameters of the supersymmetric theories, which in the
context of the minimal supersymmetric standard model (MSSM) have been
recently updated in \cite{Ali:2001jg}.  

Exclusive decays involving the $b \to s \gamma$ transition
are best exemplified by the decay $B \to K^* \gamma$, which have been 
measured with a typical accuracy of $\pm 10\%$, the current branching 
ratios being \cite{Chen:2001fj,TajimaH:2001,Aubert:2001}
${\cal B}(B^\pm \to K^{*\pm} \gamma)=(3.82 \pm 0.47) \times 10^{-5}$   
and ${\cal B}(B^0 \to K^{* 0} \gamma)=(4.44 \pm 0.35) \times 10^{-5}$.
These decays have been analyzed recently
\cite{Ali:2001ez,Beneke:2001at,Bosch:2001gv}, by taking into account 
$O(\alpha_s)$ corrections, henceforth referred to as the
next-to-leading-order (NLO) estimates,  in
the large-energy-effective-theory (LEET) limit  
\cite{Dugan:1990de,Charles:1998dr}. As this framework does not predict the
decay form  factors, which have to be supplied from outside, 
consistency of NLO-LEET estimates with current data constrains
the magnetic moment form factor in $B \to K^* \gamma$  in 
the range $T_1^{K^*}(0)=0.27 \pm 0.04$. These values are somewhat lower than 
the corresponding estimates in the lattice-QCD framework, yielding  
\cite{DelDebbio:1997kr} $T_1^{K^*}(0)=0.32^{+0.04}_{-0.02}$, and
in the light cone QCD sum rule approach, which give typically
$T_1^{K^*}(0)=0.38 \pm 0.05$ \cite{Ball:1998kk,Ali:1999mm}. (Earlier
lattice-QCD results on $B \to K^* \gamma$ form factors are reviewed in
\cite{Soni:1995qq}.) It is  imperative to check the consistency of the
NLO-LEET estimates, as this would provide a crucial test of the ideas
on QCD-factorization, formulated in the context of non-leptonic
exclusive $B$-decays~\cite{Beneke:1999br}, but which have also been
invoked in the study of exclusive rare $B$-decays 
\cite{Ali:2001ez,Beneke:2001at,Bosch:2001gv}. 

The exclusive decays $B \to K^* \ell^+ \ell^-$, $\ell^\pm =e^\pm,\mu^\pm$    
have also been studied in the NLO-LEET approach in
\cite{Beneke:2001wa,Beneke:2001at}.
 In this case, the LEET symmetry brings an enormous
simplicity, reducing the number of independent form factors from seven to
only two, corresponding to the transverse and longitudinal polarization of 
the virtual photon in the underlying process $B \to K^* \gamma^*$,
called hereafter $\xi_\perp^{(K^*)}(q^2)$ and $\xi_{||}^{(K^*)}(q^2)$. The
same symmetry reduces the number of independent form factors in the decays
$B \to \rho \ell \nu_\ell$ from four to two. Moreover, in the $q^2$-range
where the large energy limit holds, the two set of form factors are equal 
to each other, up to $SU(3)$-breaking corrections, which are already 
calculated in specific theoretical frameworks. Thus, knowing $V_{ub}$
precisely, one can make theoretically robust predictions for the 
rare $B$-decay $B \to K^* \ell^+ \ell^-$ from the measured $B \to \rho 
\ell \nu_\ell$ decay in the SM. The LEET symmetries are broken by QCD 
interactions and the leading $O(\alpha_s)$ corrections in perturbation 
theory are known \cite{Beneke:2001wa,Beneke:2001at}. 

In this talk we present the results of~\cite{Ali:2002qc}, where a systematic
analysis of the various independent helicity amplitudes in the decays
$B \to K^* \ell^+ \ell^-$ and $B \to \rho \ell \nu_\ell$ were performed 
in the NLO accuracy in the large energy limit. We recall that a  
decomposition of the final state $B \to K^* (\to K \pi) \ell^+ \ell^-$
in terms of the helicity amplitudes $H_\pm^{L,R}(q^2)$ and 
$H_0^{L,R}(q^2)$, without the explicit $O(\alpha_s)$ corrections, was
undertaken in a number of papers
\cite{Melikhov:1998cd,Aliev:1999gp,Kim:2000dq,Kim:2001xu,Nguyen:2001zu,Chen:2002bq}.
%In particular, Kim et al.~\cite{Kim:2000dq,Kim:2001xu}  
%emphasized the role of the azimuthal angle distribution as a
%precision test of the SM.
%Following closely the earlier analyses, we now
%calculate the $O(\alpha_s)$ corrections in the LEET framework.
%
Combining the analysis of the decay modes $B \to K^* \ell^+ \ell^-$
and $B \to \rho \ell \nu_\ell$, we show that 
the ratios of differential decay rates involving definite helicity states, 
$R_{-}(s)$ and $R_{0}(s)$, can be used for testing the SM precisely.
%
%----------------------------------------
\section{General framework}\hspace*{\parindent}
At the quark level, the rare semileptonic decay 
$b \to s~  \ell^+ \ell^-$ can be described in terms of the effective 
Hamiltonian  
\begin{equation}
{\cal H}_{eff} = - \frac{G_F}{\sqrt{2}}  \lambda_t  
\sum_{i=1}^{10}  C_{i}(\mu)  {\cal O}_i(\mu) \; , 
        \label{eq:he}
\end{equation}
\noi where $\lambda_t= V_{t s}^\ast  V_{tb}$ are the CKM matrix elements 
\cite{ckm} and  $G_F$ is the Fermi coupling constant. Following the
notation and the convention used in ref.~\cite{Ali:2002qc}, the above
Hamiltonian leads to the following free quark decay amplitude\footnote{
We put $m_s/m_b = 0$ and the hat denotes normalization
in terms of the $B$-meson mass, $m_B$, e.g. $\hat{s}=s/m_B^2$,
$\hat{m}_b =m_b/m_B$.}:  
\begin{eqnarray}
{\cal M} &=&  \frac{G_F \alpha_{em}}{\sqrt{2}  \pi} \lambda_t \, 
%\left
\Big\{ C_{9}  [ \bar{s}  \gamma_{\mu}  L  b ] [ \bar{\ell}  \gamma^{\mu}  \l]
%+ C_{10} [ \bar{s} \gamma_{\mu}  L  b ] [ \bar{\ell}
%\gamma^\mu  \gamma_5  \l ] 
%\right. 
\label{eq:m}\\
&&
\hspace{-1.5cm}
%\; \; \; \; \; \; \; \; \; \; \; \; \; \; 
%\; \; \; \; \; \; \; \; \; 
+ C_{10} [ \bar{s} \gamma_{\mu}  L  b ] [ \bar{\ell}
\gamma^\mu  \gamma_5  \l ]
- 2 \hat{m_{b}}  C_{7}^{\bf eff}  [ \bar{s}  i  \sigma_{\mu \nu}  {\hat{q^{\nu}}\over\hat{s}} R  b ] 
[ \bar{\l}  \gamma^{\mu}  \l ]\Big\} .\nn
\end{eqnarray} 
Here, $L/R \equiv {(1 \mp \gamma_5)}/2$, $s=q^2$, $\sigma_{\mu \nu}={i\over
2}[\gamma_{\mu},\gamma_{\nu}]$ and $q_{\mu}=(p_{+} +p_{-})_{\mu}$,
where $p_{\pm}$ are the four-momenta of the leptons. 
%Here  and in the remainder of this work we shall
%denote by $m_b \equiv m_b(\mu)$ the $\overline{MS}$ mass evaluated at a
%scale $\mu$, and by $m_{b,pole}$ the pole mass of the $b$-quark. To
%next-to-leading order the pole and $\overline{\rm MS}$ masses are related by
%%%%%%%%
%
%\begin{equation}
%\label{polerel}
%m_b(\mu) = m_{b,pole}\left(1+\frac{\alpha_s(\mu) C_F}{4\pi}\left[
%3\ln\frac{m_b^2}{\mu^2}-4\right]+O(\alpha_s^2)\right)~.
%\end{equation}
%
Since we are including the next-to-leading corrections into our
analysis, we will take the NLO $\overline{MS}$ definition of the
$b$-quark mass $m_b \equiv m_b(\mu)$ and the Wilson coefficients in next-to-leading-logarithmic order given in Table~1~ in~\cite{Ali:2002qc}.
%%%%%%%%%%%%%%%%%%%%%%%%%%%%%%%%%%%%%%%%%%%%%%%%
%
%%%%%%%%%%%%%%%%%%%%%%%%%%%%%%%%%%%%%%%%%%%%%%%%%%%%%%%%%%%%%%%%%%%%%%%%%
%\section{Form factors in the Large Energy Effective Theory}
%\hspace*{\parindent}

Exclusive $B\to V$ transitions\footnote{$V$ stands for the corresponding
vector meson.} are described by the matrix elements of the quark operators in
Eq.~(\ref{eq:m}) over meson states, which can be parameterized in
terms of the full QCD form factors (called in the literature
$A_0(q^2), A_1(q^2), A_2(q^2), V(q^2), T_1(q^2), T_2(q^2), T_3(q^2)$).

However, the factorization Ansatz enables one to relate in the
restricted kinematic region\footnote{For the $B\to K^* \ell^+ \ell^-$
decay, this region is identified as $s \leq \mbox{GeV}^2$.}, the form 
factors in full QCD  and the two corresponding LEET form factors,
namely $\xi^{(V)}_\perp(q^2)$ and $\xi^{(V)}_{||}(q^2)$ 
\cite{Beneke:2001wa,Beneke:2001at};
\begin{eqnarray}
f_k(q^2)&=& C_{\perp}\xi^{(V)}_\perp (q^2) + C_{||}\xi^{(V)}_{||}(q^2) \nn\\
&&+ \Phi_B \otimes T_k \otimes \Phi_{(V)}~,
\label{eq:fact}
\end{eqnarray}
where the quantities $C_i$ $(i=\perp, \parallel)$ encode the perturbative 
improvements of the factorized part
\begin{equation}
C_i=C_i^{(0)} + \frac{\alpha_s}{\pi} C_i^{(1)} + ... ,
\nonumber
\end{equation}
and $T_k$ is the hard spectator kernel (regulated so as to be free of 
the end-point singularities), representing the 
non-factorizable perturbative corrections, with the direct product
understood as a convolution of $T_k$ with the light-cone distribution 
amplitudes of the $B$ meson ($\Phi_B$) and the vector meson ($\Phi_V$). 
With this Ansatz, it is a straightforward exercise to implement the
$O(\alpha_s)$-improvements in the various helicity amplitudes. For
further details we refer to~\cite{Ali:2002qc}.  

%%%%%%%%%%%%%%%%%%%%%%%%%%%%%%%%%%%%%%%%%%%%%%
%%%%%%%%%%%%%%%%%%%%%%%%%%%%%%%%%%%%%%%%%%
%%
%
Lacking a complete solution of non-perturbative QCD, one has to rely on 
certain approximate methods to calculate the above form factors. We
take the ones given in \cite{Ali:1999mm}, obtained in the framework 
of Light-cone QCD sum rules. The corresponding LEET form factors
$\xi^{(K^*)}_{\perp}(s)$ and $\xi^{(K^*)}_{||}(s)$ are illustrated in
ref~.\cite{Ali:2002qc}. The range $\xi^{(K^*)}_{\perp}(s)=0.28\pm 0.04$ is
determined by the $B \rightarrow \ K^{*}\gamma$ decay rate, calculated
in the LEET approach in  
NLO order \cite{Ali:2001ez,Bosch:2001gv,Beneke:2001at} and
current data. This gives somewhat smaller values for $T_1(0)$ and $T_2(0)$
than the ones estimated with the QCD sum rules.
%
%
%%%%%%%%%%%%%%%%%%%%%%%%%%%%%%%%%%%%%%%%%%%%%%%%%%%%%%%%%%%%%%%%%%%%%%%%%%%%%%%%%%%%
\section{$B \rightarrow \ K^{*} \ell^{+} \ell^{-}$
helicity analysis}
\begin{figure}[H]
%\begin{center}
\psfrag{b}
%{\hskip -1.cm {\large{${d{\cal B}_{-} \over ds}$}} $10^{7}$}
{\hskip -0.8cm ${\cal H}_{-}^{(K^*)}$} 
\psfrag{c}{}
\psfrag{a}{}
\epsfig{file=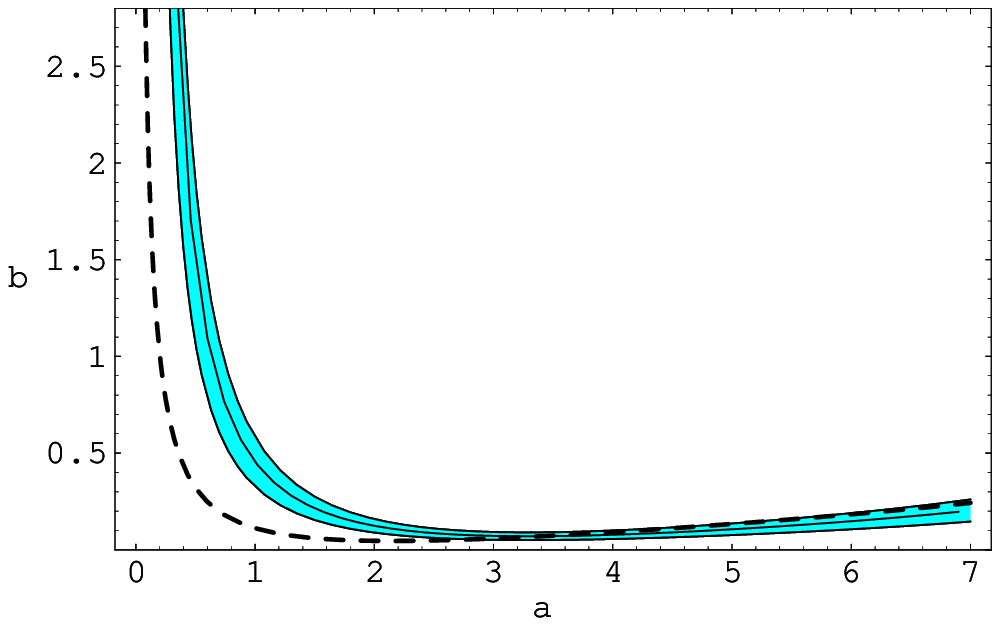,width=3.8cm,height=3cm}
%\hspace*{0.5cm}
\psfrag{b}
%{\hskip -1.5cm \large{${d{\cal B}_{|H_{+}|^2}\over ds\ 10^{-10}}$}}
{\hskip 1cm $ 10^{-3}\times{\cal H}_{+}^{(K^*)}$}
\epsfig{file=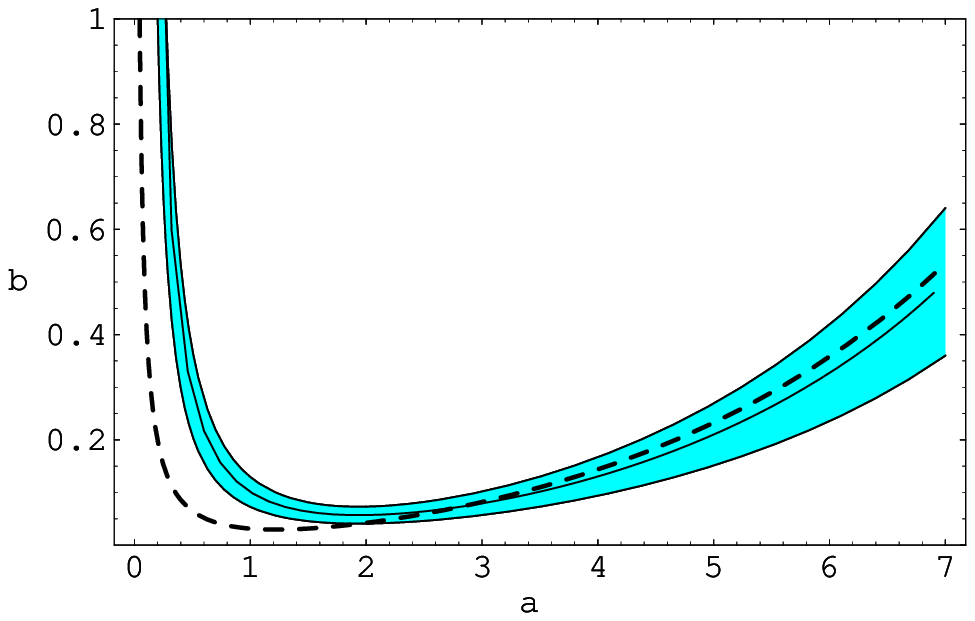,width=3.8cm,height=3cm}
%\hspace*{2cm}
\\
\psfrag{a}{\vspace*{1.cm} $\mathrm{s\ (GeV^2)}$}
\psfrag{b}
%{\hskip -1.5cm \large{${d{\cal B}_{|H_{0}|^2}\over ds\ 10^{-7}}$}}
{\hskip -0.8cm ${\cal H}_{0}^{(K^*)}$}
\epsfig{file=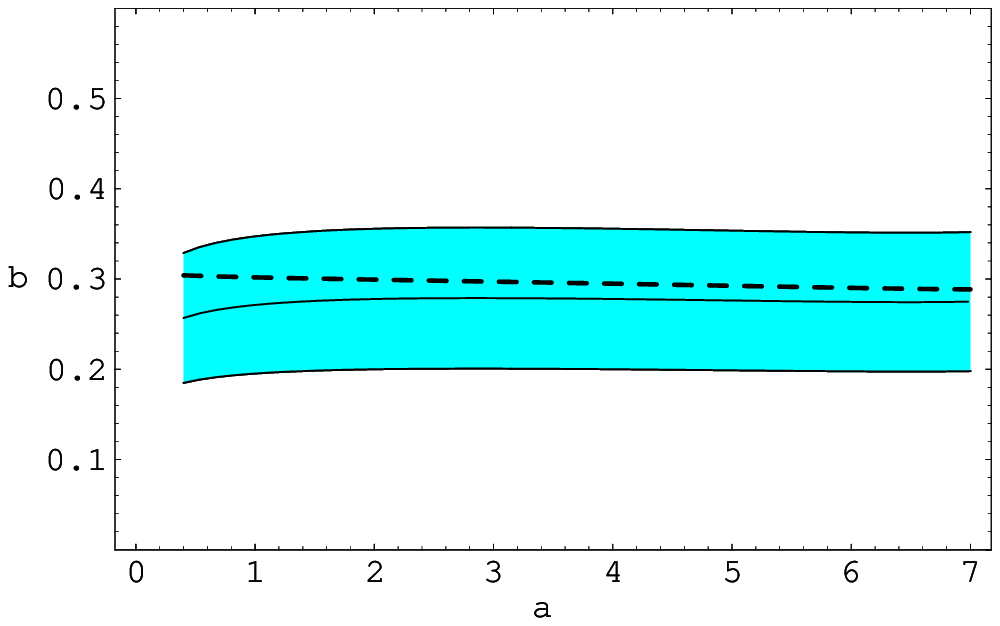,width=3.8cm,height=3cm}
%\hspace*{0.5cm}
\psfrag{b}
%{\hskip -1.5cm \large{${d{\cal B}\over ds\ 10^{-7}}$}}
{\hskip 1cm ${\cal H}^{(K^*)}$}
\psfrag{c}{}
\epsfig{file=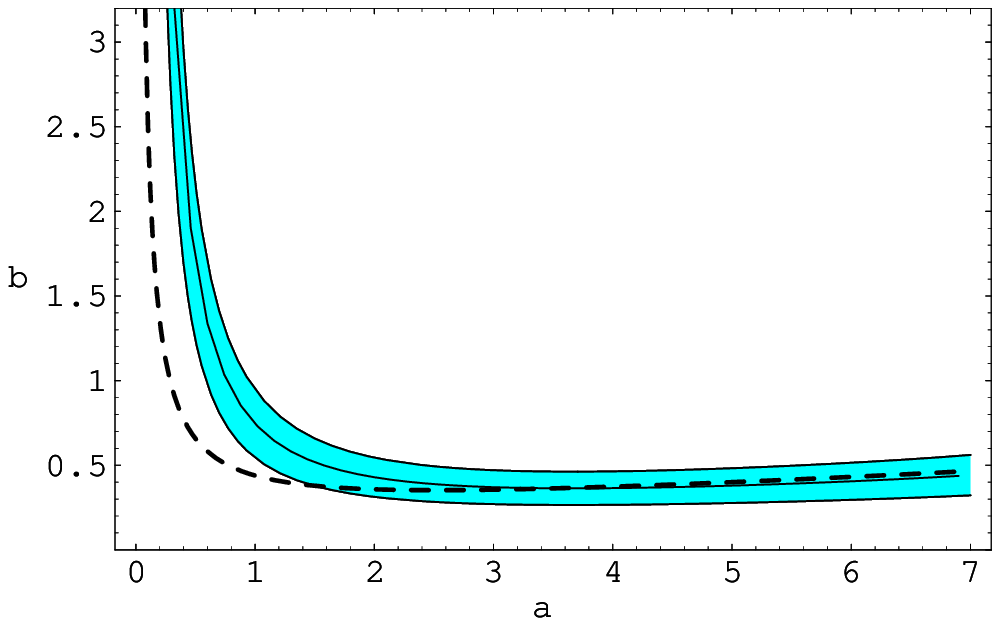,width=3.8cm,height=3cm}
\caption{\it Various individual helicity contributions $\Big({\cal
H}^{(K^*)}_{\{\pm,0\}}\equiv{d{\cal B}_{|H_{\{\pm,0\}}|^2} \over ds}
\times 10^{-7}\Big)$ and the sum $\Big({\cal H}^{(K^*)}\equiv{d{\cal B} \over
ds}\times 10^{-7}\Big)$ to 
the dilepton invariant mass distributions
for $ B\rightarrow K^*  \ell^+  \ell^-$
at NLO order (solid center line) and LO
(dashed).The band reflects the theoretical uncertainties from input
parameters.}
\label{figdBrK*}
%\end{center}
\end{figure}

\hspace*{\parindent}
Using the $B \rightarrow \ K^{*}(\to K \pi) \ell^{+} \ell^{-}$
helicity amplitudes~\cite{Kim:2000dq}, namely $H^{(K^*)}_{i}(s)~~~
(i=0,\pm 1)$ , the dilepton invariant mass
spectrum reads 
%We introduce the helicity amplitudes for the decay $B \rightarrow \
%K^{*}(\rightarrow \ K(p_K) + \pi(p_\pi )) \ell^{+}(p_+) \ell^{-}(p_-)$,
%which can be expressed as \cite{Kim:2000dq}:
%\begin{eqnarray}
%H_{\pm}^{L,R}(s)&=& (a_{L,R} \pm c_{L,R} \sqrt{\lambda})~,\nn \\
%H_{0}^{L,R}(s) &=& -a_{L,R} {(m_B^2-m_V^2 - s)\over 2~ m_{V} \sqrt{s}} +
%{b_{L,R} \lambda \over m_{V} \sqrt{s}}~,\label{Hs}
%\end{eqnarray}
% 
%\noi where $ \lambda=\left[{1\over 4}(m_{B}^2-m_{V}^2-s)^2 - m_{V}^2\
%s\right]$ and the quantities 
%$a_{L,R}$, $b_{L,R}$ and $c_{L,R}$ can be found in~\cite{Ali:2002qc}.
%Using the above helicity amplitudes,  the dilepton invariant mass
%spectrum reads
% 
\begin{eqnarray} 
{d{\cal B} \over ds} &=& \tau_B {\alpha_{em}^2 G_{F}^2\over 384 \pi^5} 
 \sqrt{\lambda} {m_{b}^2\over m_{B}^3 }  \lambda_t^2
\sum_{i=0,\pm1}|H^{(K^*)}_{i}(s)|^2,\nn\\ 
% \Big\{|H_{+}(s)|^2 + 
% |H_{-}(s)|^2+|H_{0}(s)|^2 \Big\}~,\label{dBrdl2}\\ 
 &=& {d{\cal B}_{|H_{+}|^2} \over ds}+{d{\cal B}_{|H_{-}|^2} \over ds}+{d{\cal B}_{|H_{0}|^2} \over ds}.
\end{eqnarray}
where $ \lambda=\left[{1\over 4}(m_{B}^2-m_{V}^2-s)^2 - m_{V}^2\
s\right]$. Using the input parameters presented in~\cite{Ali:2002qc},
we have plotted in Fig.~(\ref{figdBrK*}), the dilepton invariant mass
spectrum $d{\cal B}_{|H_{\{0,\pm\}}|^2}/ ds$ and the total dilepton
invariant mass, showing in each case the leading order and the 
next-to-leading order results. 
\vspace{-0.45cm}
As can be seen from Fig.~(\ref{figdBrK*}) the total decay rate is
dominated by the contribution from the helicity $|H_{-}|$ component,
whereas the contribution proportional to the helicity amplitude
$H_{+}(s)$ is negligible.
The next-to-leading order correction to the lepton invariant mass
spectrum in $ B \rightarrow K^* \ell^+ \ell^-$ is significant in the 
low dilepton mass region ($s \leq 2$ GeV$^2$), but small beyond that
shown for the anticipated validity of the LEET theory ($s \leq 8$
GeV$^2$). Theoretical uncertainty in our prediction is mainly due to
the form factors, and to a lesser extent due to the
parameters $\lambda_{B,+}^{-1}$ and the $B$-decay
constant, $f_B$.
%%%%%%%%%%%%%%%%%%%%%%%%%%%%%%%%%%%%%%%%%%%%%%%%%%%%%%%%%%%%%%%%%%%%%%%%
%\subsection{\bf Forward-Backward asymmetry}
%\hspace*{\parindent}

\begin{figure}[H]
\psfrag{a}{$s\ (GeV^2)$}
\psfrag{b}{\hskip -2.5cm $dA_{FB}/ ds$}
\psfrag{c}{\hskip 0cm }
%\includetext{vighvgvcdigvvwqljcdhvuqvlwqhjcvwqglv}
\begin{center}
\includegraphics[width=8cm,height=4cm]{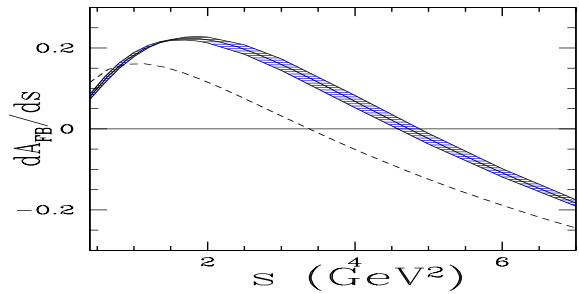}
\caption{\it 
FB-asymmetry at NLO order (solid center line)
and LO (dashed). The band reflects the theoretical uncertainties from
the input parameters.}
\label{FigFBA}
\end{center}
\end{figure}

Besides the differential branching ratio, $B \to K^* \ell^+ \ell^-$
decay offers other distributions (with different combinations of
Wilson coefficients) to be measured.
An interesting quantity is the Forward-Backward (FB) asymmetry
defined in \cite{Ali:1991is}
%The differential forward-backward asymmetry (FBA) is defined as\cite{Ali:1991is}
\begin{equation}
  \frac{\d \a_{\rm FB}}{\d \sh} \equiv 
        -\int_0^{\uh(\sh)} \d\uh \frac{\d^2\gl}{\d\uh~ \d\sh}
              + \int_{-\uh(\sh)}^0 \d\uh \frac{\d^2\gl}{\d\uh~ \d\sh} \; .
  \label{eq:dfba}
\end{equation}
Where the kinematic variable $\hat{u}  \equiv  (\hat{p}_{B} -
\hat{p}_-)^2 - (\hat{p}_{B} -\hat{p}_+)^2$ \footnote{which are bounded as 
$ (2 \hat{m}_{\ell})^2 \leq  \hat{s}  \leq (1 - \hat{m}_{K^*})^2$,
$ -\hat{u}(\hat{s}) \leq  \hat{u}  \leq \hat{u}(\hat{s})$,
with $\hat{m}_{\ell}=m_{\ell}/m_{B}$ and
$\hat{u}(\hat{s})={2\over m_{B}^2}
\sqrt{\lambda(1-4{\hat{m}_{l}^2\over\hat{s}})}$.}.   
Our results for FBA are shown in Fig.~\ref{FigFBA} in the  LO and NLO
accuracy.
%%%%%%%%%%%%%%%%%%%%%%%%%%%%
%\hspace*{2cm}
\vspace{-0.45cm}
At the LO the location of the FB-asymmetry zero is
$s_0\simeq3.4\, \mbox{GeV}^2$, which is substantially shifted at the
NLO by $\sim 1\, \mbox{GeV}^2$ leading to $s_0\simeq 4.7\, \mbox{GeV}^2$.
We essentially confirm the results obtained in the NLO-LEET
context by {\it Beneke et al.} \cite{Beneke:2001at}.
%%%%%%%%%%%%%%%%%%%%%%%%%%%%%%%%%%%%%%%%%%%%%%%%%%%%%%%%%%%%%%%%%%%%%%%%%%%%%%
%\newpage \pagestyle{plain}
\section{$B \rightarrow \rho \ell \nu_{\ell}$ helicity analysis}
\begin{figure}[H]
\psfrag{b}
{\hskip -0.5cm ${\cal H}_{0}^{(\rho)}$}
%{\hskip 1.5cm \large{${d{\cal B}_{|H_{-}|^2}\over ds\|V_{ub}|^2 }$}} 
\psfrag{c}{}
\psfrag{a}{}
\begin{center}
\includegraphics[width=3.8cm,height=3cm]{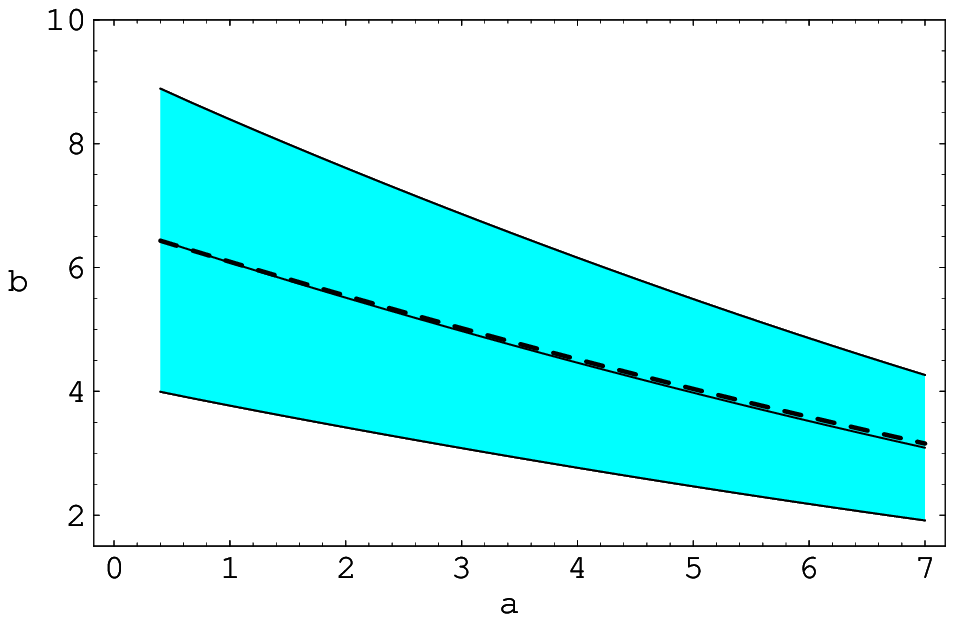}
%\hspace*{0.3cm}
\psfrag{b}
%{\hskip 1cm \large{${d{\cal B}_{|H_{+}|^2}\over ds\ |V_{ub}|^2 10^{-4} }$}}
{\hskip 0.8cm $10^{-4}\times {\cal H}_{+}^{(\rho)}$}
\includegraphics[width=3.8cm,height=3cm]{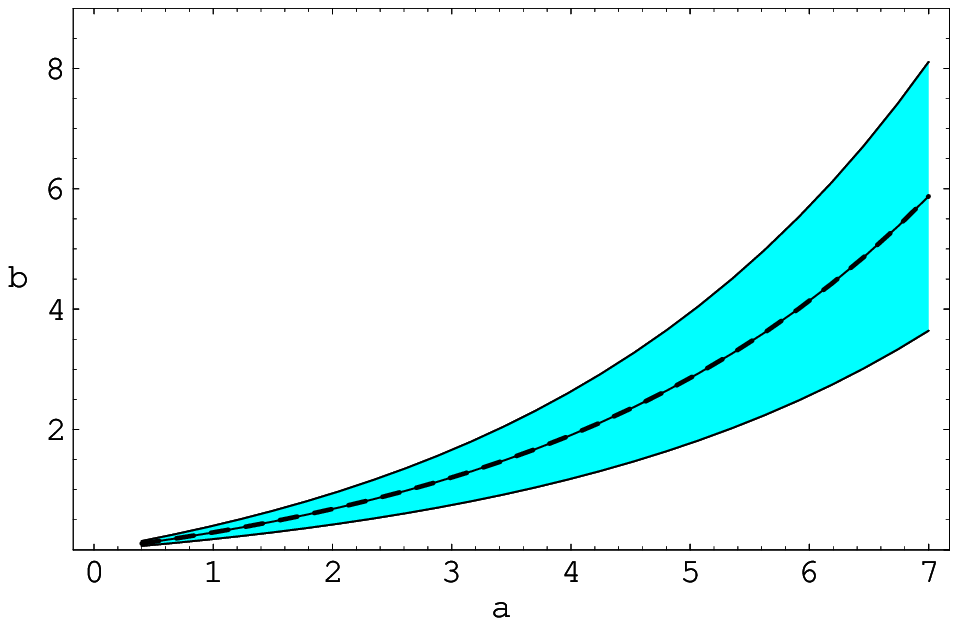}
\\
\psfrag{a}{\hskip 0.cm $\mathrm{s\ (GeV^2)}$}
\psfrag{b}
%{\hskip 1.5cm \large{${d{\cal B}_{|H_{0}|^2}\over ds\|V_{ub}|^2 }$}}
{\hskip -0.5cm ${\cal H}^{(\rho)}$}
\includegraphics[width=3.8cm,height=3cm]{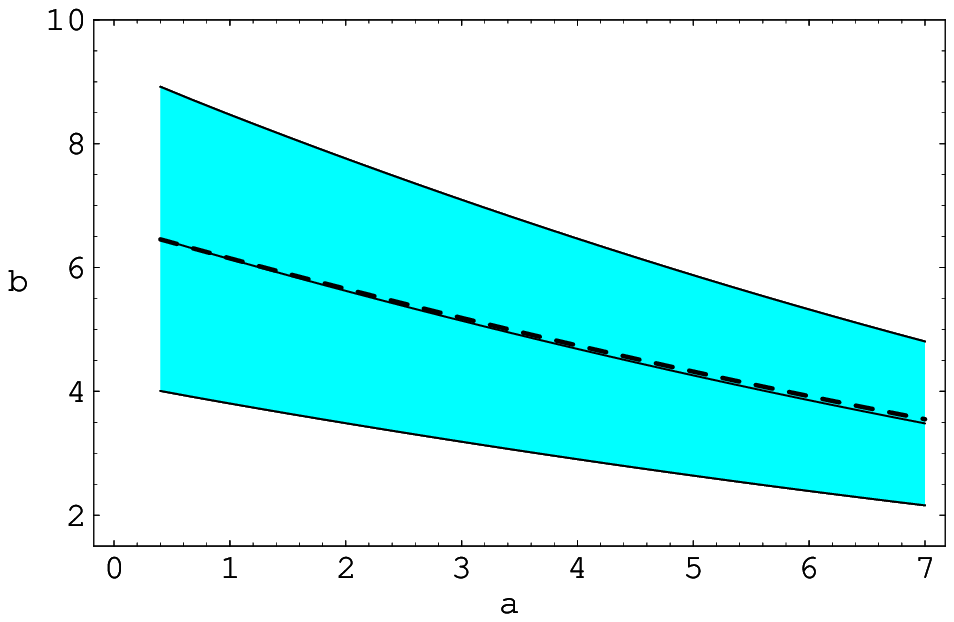}
%\hspace*{0.5cm}
\psfrag{b}
%{\hskip 1.5cm \large{${d{\cal B}\over ds\ |V_{ub}|^2 }$}}
{\hskip 0.8cm ${\cal H}_{-}^{(\rho)}$}
\includegraphics[width=3.8cm,height=3cm]{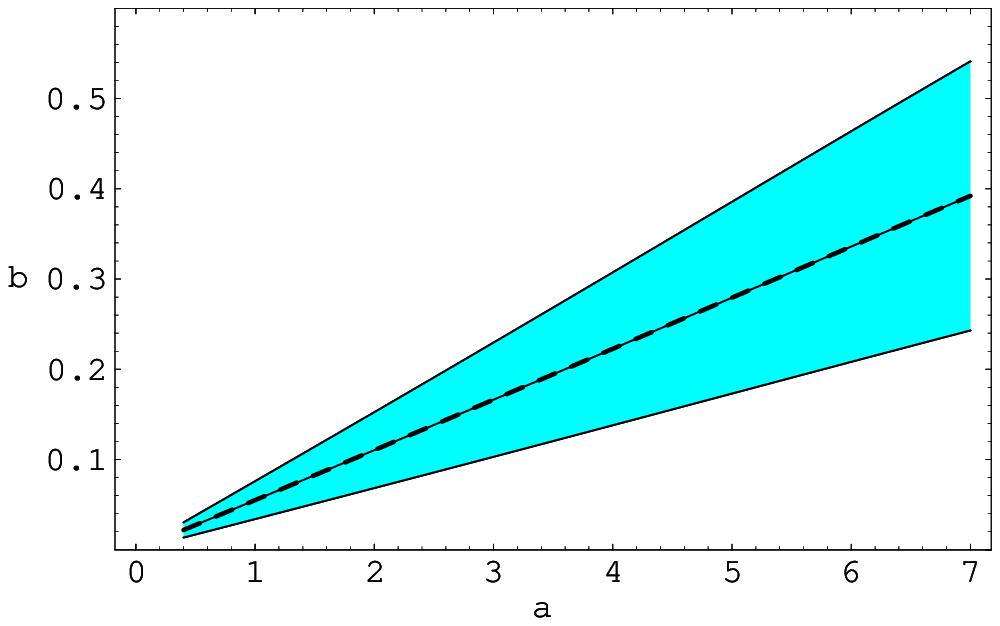}
\caption{\it Various individual helicity contributions $\Big({\cal
H}^{(\rho)}_{\{\pm,0\}}\equiv{d{\cal B}_{|H_{\{\pm,0\}}|^2} \over ds\
|V_{ub}|^2}\Big)$ and the sum $\Big({\cal H}^{(\rho)}\equiv{d{\cal B }
\over ds|V_{ub}|^2}\Big)$ to
dilepton invariant mass
distributions for $B \rightarrow \rho \ell \nu_{\ell}$ at
NLO order (solid center line) and LO (dashed).The band reflects the
theoretical uncertainties from input parameters.} 
\label{dBHrho}
\end{center}
\end{figure}

\hspace*{\parindent}
%\noi The differential decay rate for $B\rightarrow \rho (\rightarrow
%\pi^{+} \pi^{-}) \ell \bar{\nu_{\ell}}$ can be  expressed as follows
%\cite{Korner:1989qb,Korner:1989tb,Richman:wm}:
%
The helicity  amplitudes for $B\rightarrow \rho (\rightarrow
\pi^{+} \pi^{-}) \ell \nu_{\ell}$, namely
$H^{(\rho)}_{i}(s)~~~(i=0,\pm 1)$,  can in turn be related to the two
axial-vector form factors, $A_{1}(s)$ and $A_{2}(s)$, and the vector
form factor, $V(s)$, which appear in the hadronic current 
\cite{Richman:wm}.
The $B\rightarrow \rho (\rightarrow \pi^{+} \pi^{-}) \ell
\nu_{\ell}$ total branching decay rate\footnote{we put ${\cal B}(\rho \rightarrow \pi^+ \pi^-)=100\%.$} can be expressed in terms
of the corresponding helicity amplitudes as~\cite{Korner:1989qb,Richman:wm}
\begin{eqnarray}
{d{\cal B} \over ds} &=& \tau_B {G_F^2\ s\ \sqrt{\lambda} \over 96
m_{B}^3 \pi^4} |V_{ub}|^2 
%({\it{{\cal B}(\rho \rightarrow \pi^+ \pi^-)}}) 
\sum_{i=0,\pm1}|H^{(\rho)}_{i}(s)|^2,\nn\\ 
%\Big\{|H_0(s)|^2 +|H_{+}(s)|^2 + |H_{-}(s)|^2 \Big\} \label{dgamads}\\ 
{} &=& {d{\cal B}_{|H_{0}|^2} \over ds}+{d{\cal B}_{|H_{+}|^2} \over ds}+{d{\cal B}_{|H_{-}|^2}
 \over ds} ~.\label{dgamads2}
\end{eqnarray}
The contributions from the $\vert H^{(\rho)}_{-}(s)\vert^2$, $\vert
H^{(\rho)}_{+}(s)\vert^2$, $\vert H^{(\rho)}_{0}(s)\vert^2$ and the total are shown in
Fig.~(\ref{dBHrho}). Contrary to
the $B \rightarrow K^* \ell^+ \ell^- $ decay rate, the
$B \rightarrow \rho \ell \nu_\ell$
 decay is dominated by the helicity-0
component. The impact of the NLO
correction on the various branching ratios in $B \rightarrow \rho \ell
\nu_\ell$ is less significant than in the $B \rightarrow K^* \ell^+
\ell^- $ decay,  reflecting the absence of the penguin-based amplitudes
in the former decay. 
\vspace{-0.45cm}
Concerning the $B \rightarrow \rho \ell \nu_\ell$ form factors,
one has to consider the SU(3)-breaking effects in relating them to the
corresponding form factors in $B \to K^* \ell^+ \ell^-$. For the form
factors in full QCD, they have been evaluated within the QCD sum-rules
\cite{Ali:vd}. These can be taken to hold also for the ratio of the LEET
form factors. Thus, we take
\begin{equation}
\xi^{(\rho)}_{\perp,||}(s)={\xi^{(K^*)}_{\perp,||}(s) \over
\zeta_{SU(3)}}~.
 \label{eq:paraball}
\end{equation}
While admitting that this is a somewhat simplified picture, as the
effect of $SU(3)$-breaking is also present in the $s$-dependent
functions, but checking numerically the functions resulting from
Eq.~(\ref{eq:paraball}) with the ones worked out for the full QCD form
factors in the QCD sum-rule approach in \cite{Ball:1998kk}, we find that
the two descriptions are rather close numerically in the region of interest
of $s$.

%%%%%%%%%%%%%%%%%%%%%%%%%%%%%%%%%%%%%%%%%%%%%%%%%%%%%%%%%%%%%%%%%%%%%%%%%%%
\section{Determination of $|V_{ub}|/|V_{ts}|$ 
%from $B \to \rho \ell
%\nu_\ell$ and\\ $B \to K^* \ell^+ \ell^-$ Decays
}
\begin{figure}[t]
\begin{center}
\includegraphics[width=3.8cm,height=3cm]{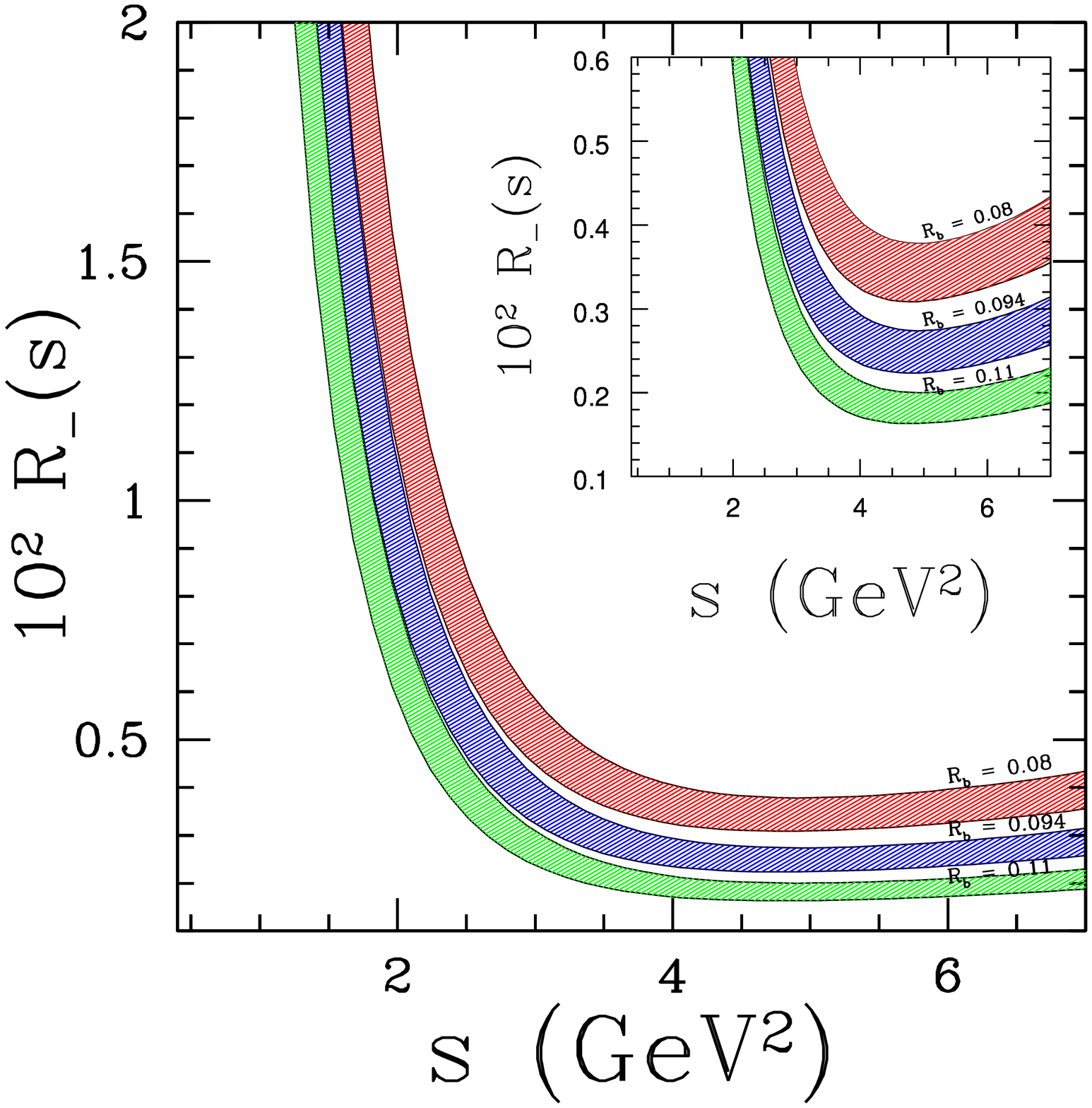}
%\hspace*{0.5cm}
\includegraphics[width=3.8cm,height=3cm]{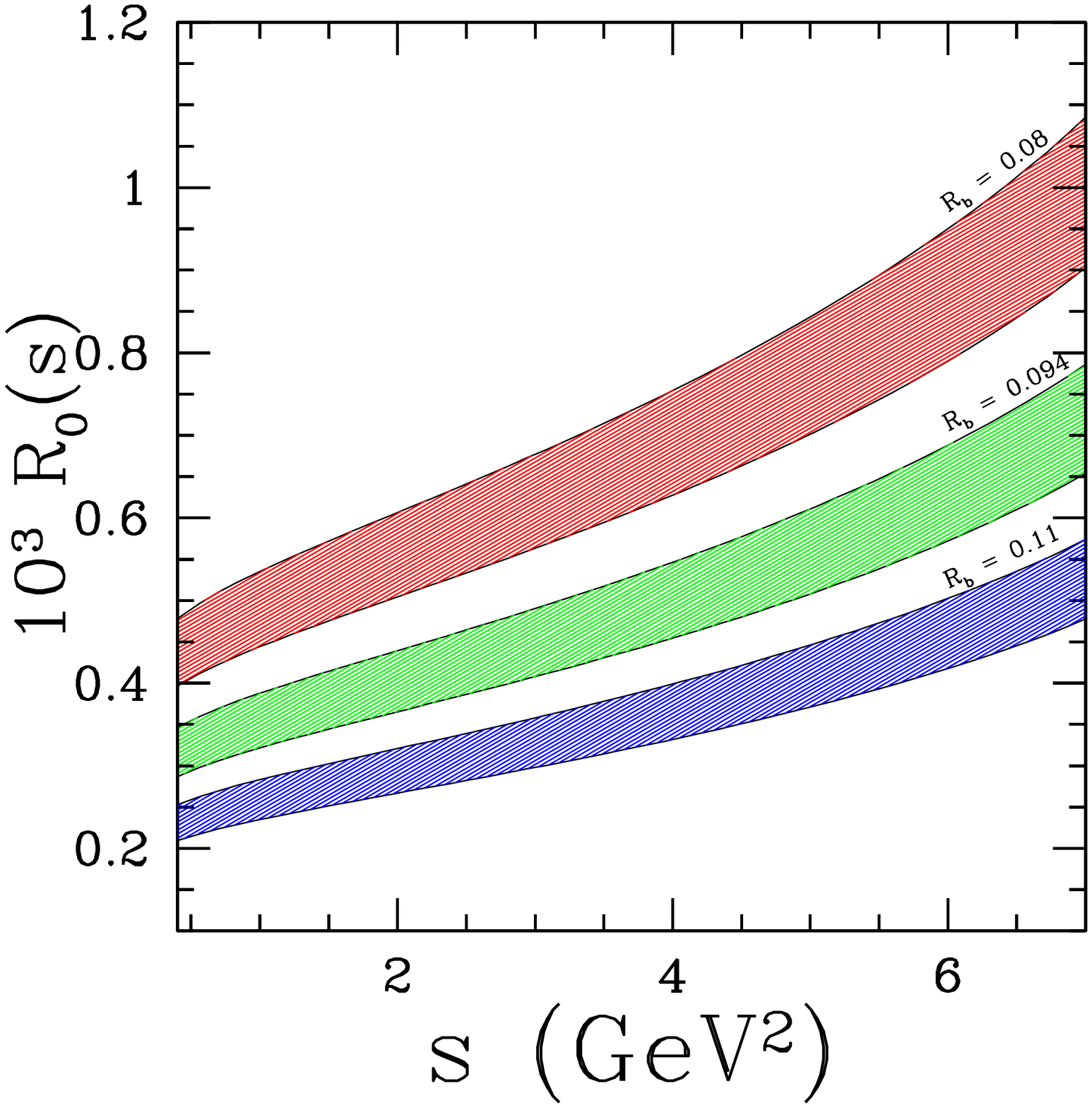}
\caption{\it The Ratios $R_-(s)$ (left-hand plot) and $R_0(s)$
(right-hand plot)  with three indicated values of the CKM 
ratio $R_{b} \equiv |V_{ub}|/|V_{tb}V_{ts}^*|$. The
bands reflect the theoretical uncertainty from $\zeta_{SU(3)}=1.3 \pm
0.06$  and $\xi^{(K^*)}_{{\perp}}(0)=0.28\pm 0.04$.}
\label{RVub}
\end{center}
\end{figure}

\hspace*{\parindent}
The measurement of exclusive $B \rightarrow \rho  \ell \nu_{\ell}$ decays
is one of the major goals of B physics. It provides a good tool for
the extraction of  $|V_{ub}|$, provided the form factors can be either
measured precisely or calculated from first principles, such as the
lattice-QCD framework.  To reduce the non-perturbative uncertainty in the
extraction of $V_{ub}$, we propose to study the ratios of the
differential decay rates in $B \to \rho \ell \nu_\ell$ and $B \to K^* \ell^+
\ell^-$ involving definite helicity states.  These $s$-dependent ratios 
$R_{i}(s)$, $(i=0,-1,+1)$ are defined as follows: 
\begin{eqnarray}
R_{i}(s) ={{d\Gamma_{H_i}^{B \rightarrow  K^{*} \ \ell^{+}  \ell^{-}}/ds}
\over{d\Gamma_{H_i}^{B \rightarrow  \rho \ \ell \nu_{\ell}}/ds}}
\label{Ratio} 
\end{eqnarray}
The ratio $R_{-}(s)$ suggests itself as the most interesting one, as the
form factor dependence essentially cancels. From this, one can measure
the ratio $\vert V_{ts} \vert/\vert V_{ub}\vert$. In
Fig.~\ref{RVub}, we plot $R_{-}(s)$ and $R_{0}(s)$ for three
representative values of the 
CKM ratio $R_b = \vert V_{ub}\vert/\vert V_{tb} V_{ts}^* \vert =
\vert V_{ub}\vert/\vert V_{cb}\vert =0.08$, $0.094$, and $0.11$. 
%%%%%%%%%%%%%%%%%%%%%%%%%%%%%%%%%%%%
However, as we
remarked earlier, the ratio $R_{-}(s)$ may be statistically limited due to
the dominance of the decay $B \to \rho \ell \nu_\ell$ by the Helicity-$0$
component. Hence, we also show the ratio $R_{0}(s)$, where the
form factor dependence does not cancel. For the LEET form factors used here,
the compounded theoretical uncertainty is shown by the shaded regions.  
This figure suggests that high statistics experiments may be able to
determine the CKM-ratio from measuring $R_{0}(s)$ at a competitive level
compared to the other methods {\it en vogue} in experimental studies.
%\vspace*{-1cm}
%%%%%%%%%%%%%%%%%%%%%%%%%%%%%%%%%%%%%%%%%%%%%%%%%%%%%%%%%%%%%%%%%%%
\section{
%The Ratios $R_-(s)$ and $R_0(s)$ as Probes of
 SUSY effect in $B \rightarrow  K^{*} \ \ell^{+}  \ell^{-}$}
\hspace*{\parindent}
In order to look for new physics in $B \rightarrow  K^{*} \ell^{+}
\ell^{-}$,  we propose to study  the ratio $R_{\{0,-\}}(s)$,
introduced in Eq.~(\ref{Ratio}). To illustrate generic SUSY effects
in ${B \rightarrow  K^{*} \ell^{+} \ell^{-}}$, 
we note that the Wilson coefficients $C^{\bf eff}_7$, $C^{\bf eff}_8$,
$C_9$ and $C_{10}$ receive additional contributions from the supersymmetric
particles. We incorporate these effects by assuming that the ratios of the
Wilson coefficients in these theories and the SM deviate from 1. These ratios
are defined as follows:
\begin{eqnarray}
r_{k}(\mu) = {C_{k}^{SUSY}\over C_{k}^{SM}},~~~ (k= 7, \cdots, 10).
\label{rk} 
\end{eqnarray}
\noi They depend on the renormalization scale (except for $C_{10}$),
for which we take $\mu =m_{b,pole}$. For the sake of illustration,
we use representative values for the large(small)-$\tan \beta$  SUGRA model,
in which the ratios $r_7$ and $r_8$ actually change (keep) their signs. The
supersymmetric effects on the other two Wilson coefficients $C_9$ and
$C_{10}$ are generally small in the SUGRA models, leaving $r_9$ and $r_{10}$
practically unchanged from their SM value. To be specific, we take 
\footnote{We thank Enrico Lunghi for providing us with these numbers.}
$r_{7} = -1.2,\ \ r_{8} = -1,\ \    \ r_{9} = 1.03,\ \ r_{10} = 1.0~$
($r_{7} = 1.1,\ \ r_{8} = 1.4,\ \   \ r_{9} = 1.03,\ \ r_{10} = 1.0$).
\begin{figure}[t]
\begin{center}
\includegraphics[width=3.8cm,height=3cm]{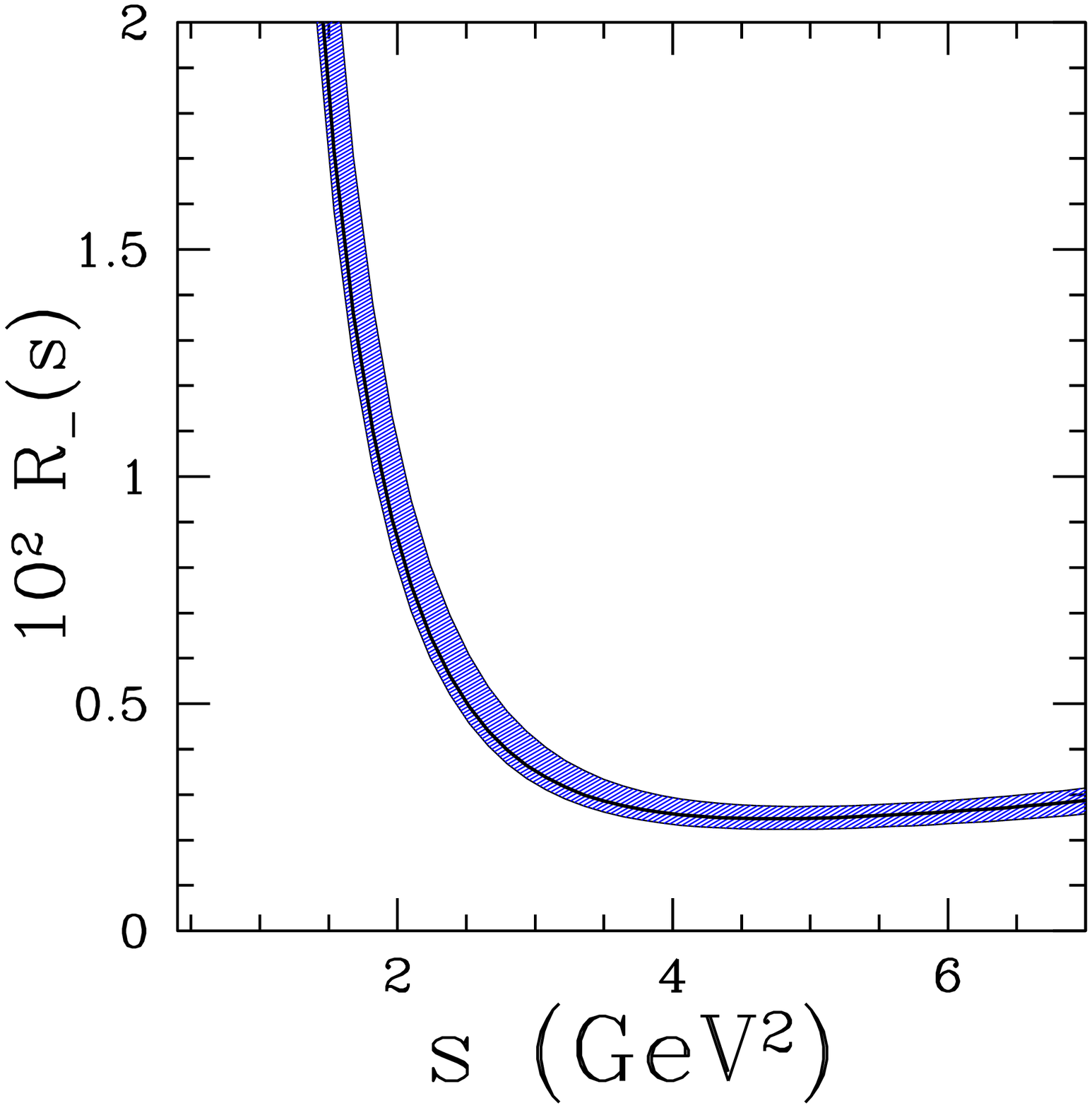}
%\hspace*{1cm}
\includegraphics[width=3.8cm,height=3cm]{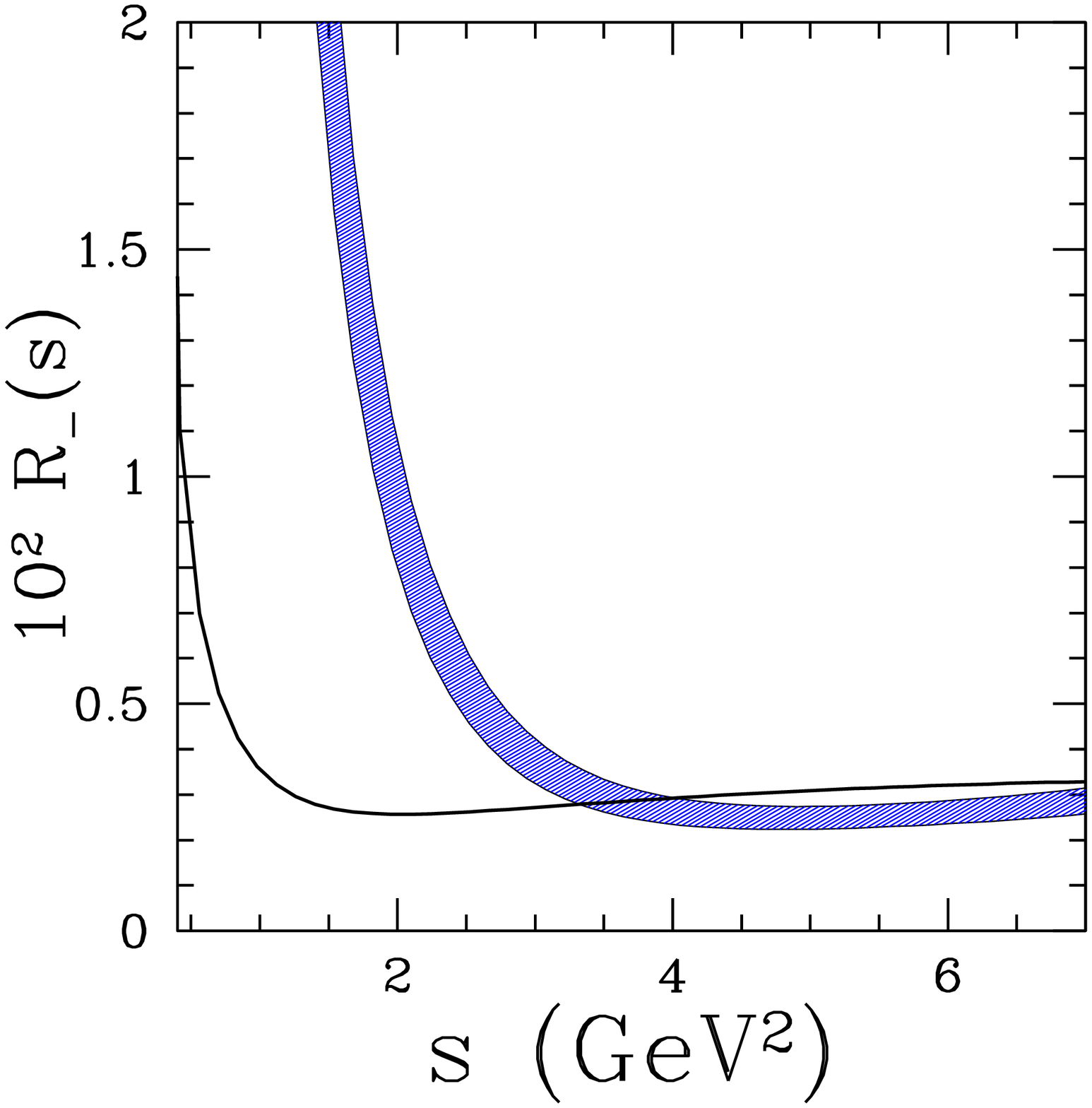}
\\
\includegraphics[width=3.8cm,height=3cm]{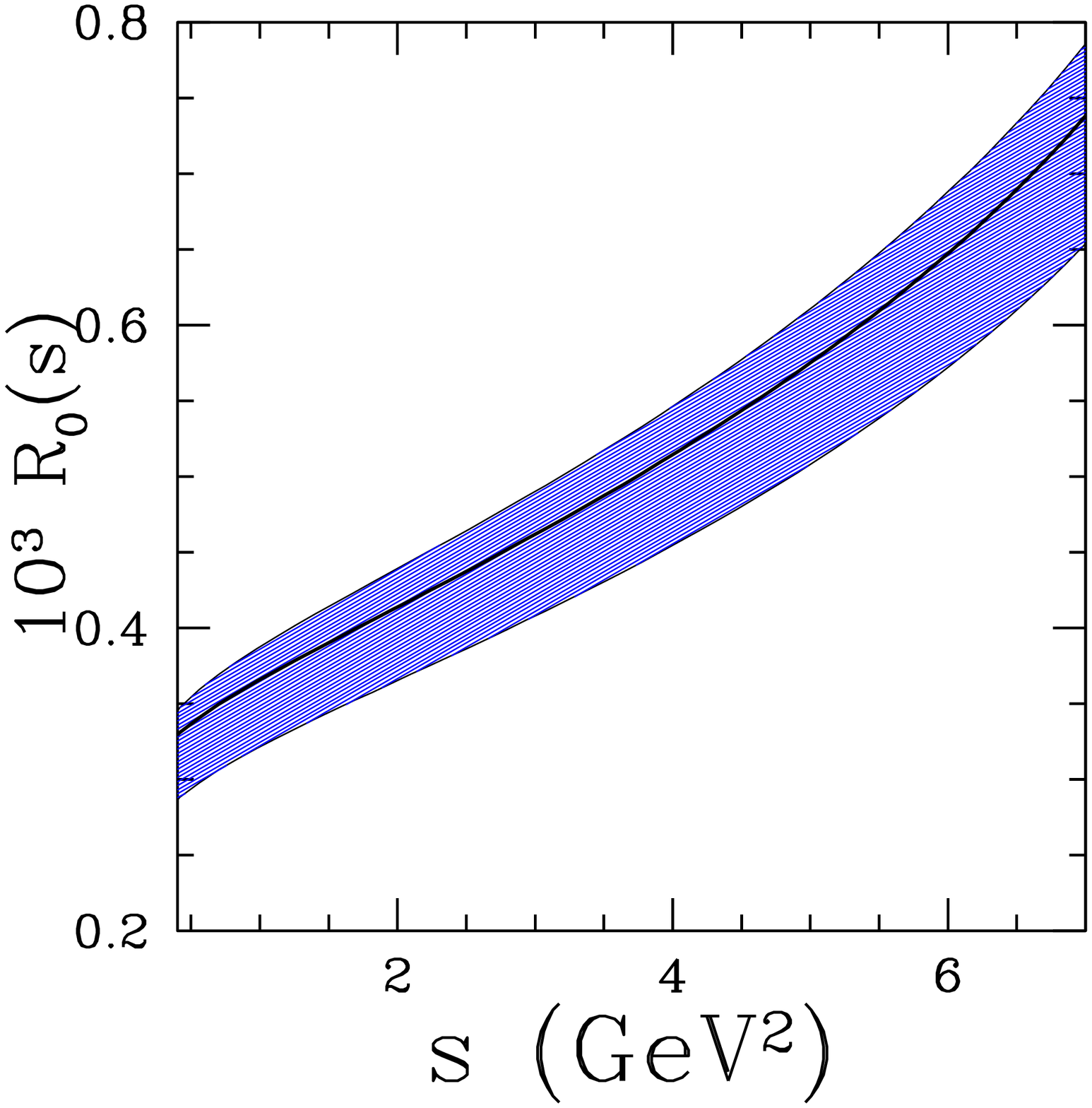}
%\hspace*{1cm}
\includegraphics[width=3.8cm,height=3cm]{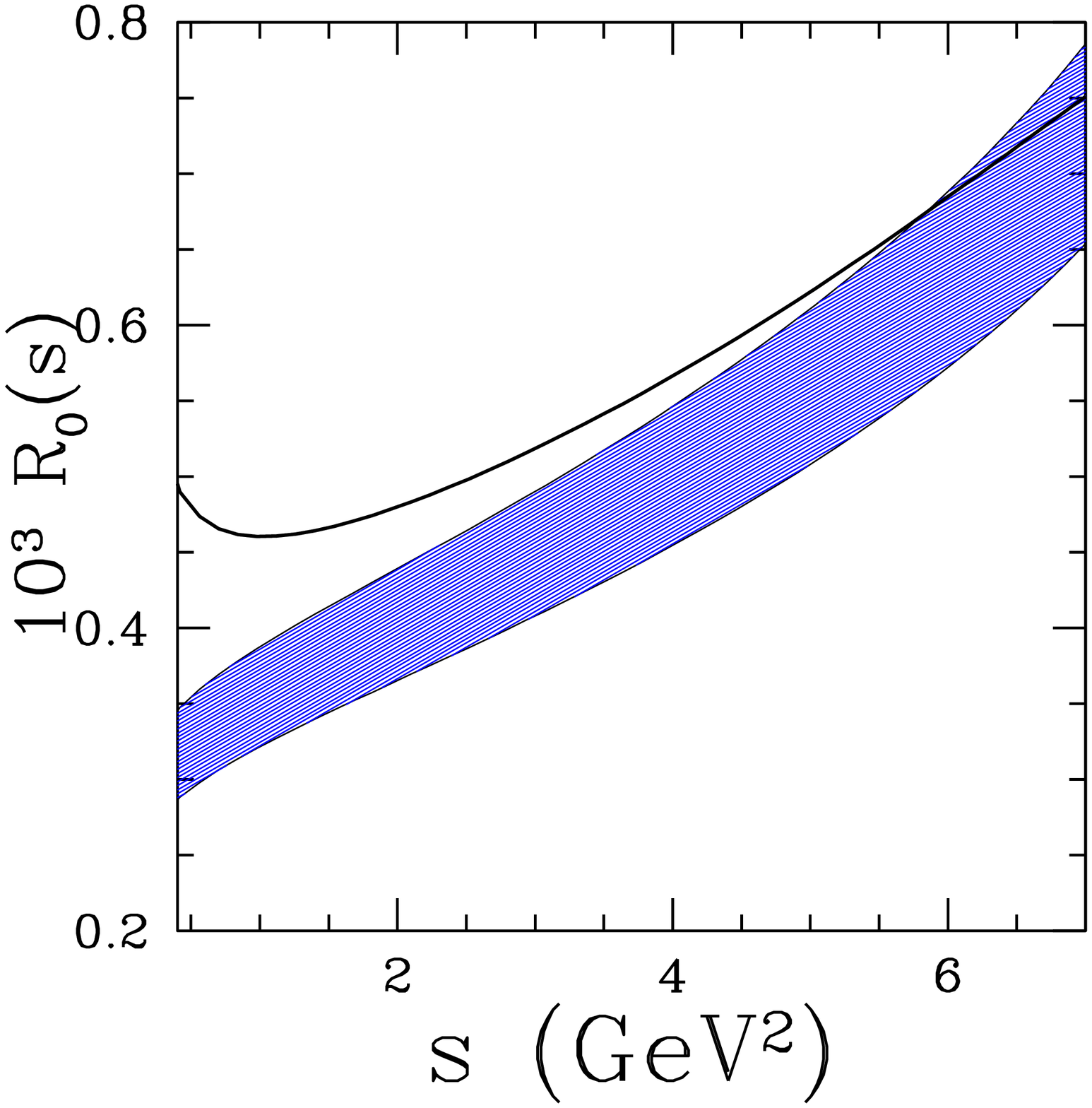}
\caption{\it 
The Ratio $R_{\{-,0\}}(s)$ in the SM with
$R_b=0.094$, $\zeta_{SU(3)}=1.3$, 
$\xi^{(K^*)}_{{\perp}}(0)=0.28$ and in SUGRA, with $(r_7,\ r_8)=(1.1,\
1.4)$ (left-hand plots) and  $(r_7,\ r_8)=( -1.2,\ -1)$ (right-hand plots).
The SM and the SUGRA contributions are  represented respectively by the shaded
area and the solid curve. The shaded area depicts the theoretical
uncertainty on $\zeta_{SU(3)}=1.3 \pm 0.06$  and on
$\xi^{(K^*)}_{{\perp}}(0)=0.28\pm 0.04$.}
\label{RSMetSUGRA}
\end{center}
\end{figure}
In Fig.~\ref{RSMetSUGRA}, we present a
comparative study of the SM and SUGRA partial distribution 
for $H_{-}$ and $H_{0}$. In doing this, we also show the
attendant theoretical uncertainties for the SM, worked out in the LEET
approach. For
these distributions, we have used the form factors from 
\cite{Ali:1999mm} with the SU(3)-symmetry breaking parameter taken in the
range $\zeta_{SU(3)}=1.3\pm 0.06$.

From Fig.~\ref{RSMetSUGRA}-{\it left-hand plots}, where
$r_{k}>0$, it is difficult to work out a signal of new
physics from the SM picture. There is no surprise to be expected, due to the
fact that in these scenario the corresponding ratio $r_k$ is
approximatively one, which makes the SUGRA picture almost the same as
in the SM one. 
However, Fig.~\ref{RSMetSUGRA}-{\it right-hand plots} with
$(r_7,\ r_8)<0$ illustrates clearly that despite non-perturbative
uncertainties, it is possible, in principle,  in the low $s$ region  to
distinguish between the SM and a SUGRA-type models, provided the ratios
$r_k$ differ sufficiently from 1. 
%
%%%%%%%%%%%%%%%%%%%%%%%
%
%\vskip 5mm %------------------------------------------------------
\section{Summary}
\hspace*{\parindent}
In this talk,  we have reported an
$O(\alpha_s)$-improved analysis of the various helicity components
in the decays $B \to K^* \ell^+ \ell^-$ and $B \to \rho \ell \nu_\ell$,
carried out in the context of the Large-Energy-Effective-Theory.
%The underlying symmetries in the large energy limit lead to an enormous
%simplification as they reduce the number of independent form factors
%in these decays. The LEET-symmetries are broken by QCD
%corrections, and we have calculated the helicity components
%implementing the $O(\alpha_s)$ corrections. 
The results presented here make use of the form factors calculated in
the QCD sum rule approach. The 
LEET form factor $\xi_\perp^{(K^*)}(0)$ is constrained by current data on $B
\to K^* \gamma$. As the theoretical analysis is restricted to the lower
part of the dilepton invariant mass region in $B \to K^* \ell^+ \ell^-$,
typically $s \leq 8$ GeV$^2$, errors in this form factor are
not expected to severely limit theoretical precision. This implies that
distributions involving the $H_{-}(s)$ helicity component can be
calculated reliably. Precise measurements of the two LEET form factors
$\xi_{\perp}^{(\rho)}(s)$ and $\xi_{\parallel}^{(\rho)}(s)$ in
the decays $B \to \rho \ell \nu_\ell$ can be used to largely reduce the
residual model dependence. With the assumed form factors, we have worked
out a number of single (and total) distributions in $B \to \rho
\ell \nu_\ell$, which need to be confronted with data. 
We also show the
$O(\alpha_s)$ effects on the FB-asymmetry, confirming
essentially the result found in~\cite{Beneke:2001at}. Combining the
analysis of the decay modes  
$B \to K^* \ell^+ \ell^-$ and $B \to \rho \ell \nu_\ell$, we show that
the ratios of differential decay rates involving definite helicity states, 
$R_{-}(s)$ and $R_{0}(s)$, can be used for testing the SM precisely.
We work out the dependence of these ratios on the CKM matrix elements
$\vert V_{ub}\vert/\vert V_{ts}\vert$.
 
We have also analyzed possible effects on these
ratios from New Physics contributions, exemplified by representative
values for the effective Wilson coefficients in SUGRA models. 
%The main thrust of this paper lies, however, on showing that the currently
% prevailing theoretical uncertainties on the SM distributions in $B
% \to K^*\ell^+ \ell^-$ can be largely reduced by using the LEET approach and data on $B \to K^* \gamma$ and $B \to \rho \ell \nu_\ell$ decays. 
Finally, we remark that the current experimental limits on $B \to
K^* \ell^+ \ell^-$ decay (and the observed $B \to X_s \ell^+ \ell^-$
 and $B \to K \ell^+ \ell^-$ decays)
\cite{bellebsll,babarbsll,Affolder:1999eb,Anderson:2001nt} 
are already probing the SM-sensitivity. With the integrated luminosities   
over the next couple of years at the $B$ factories, the
helicity analysis in  $B \to \rho \ell \nu_\ell$
and $B \to K^* \ell^+ \ell^-$ decays presented here can be
carried out experimentally.
%%%%%%%%%%%%%%%%%%%%%%%%%%%%%%%%%%%%%%%%%%%%%%%%%%%%%%%%%%%%%%%%%%%%%%%%%
%%%%%%%%%%%%%%%%%%%%%%%%%%%%%%%%%%%%%%%%%%%%%%
\section*{Acknowledgments}
\hspace*{\parindent} 
It is a great pleasure to thank Ahmed Ali for the fruitful collaboration and
useful remarks on the manuscript. For the work presented here, I also
gratefully acknowledge DESY for financial support.

%%%%%%%%%%%%%%%%%%%%%%%%%%%%%%%%%%%%%%%%%%%%%%%%

\end{document}